\documentclass[10pt, conference]{IEEEtran}

\usepackage{amsmath}
\usepackage{graphicx}
\usepackage{epstopdf}
\usepackage{amssymb}
\usepackage{amsthm}
\usepackage{caption}
\usepackage{cite}
\usepackage{subfigure}
\usepackage{enumerate}

\begin{document}
\title{Power Control Factor Selection in Uplink OFDMA Cellular Networks}
\author{
\authorblockN{Suman Kumar and  K. Giridhar }
\authorblockA{Dept.\ of Electrical Engineering\\
Indian Institute of Technology Madras \\
Chennai, 600036, India\\
Email: \{suman, giri\}@tenet.res.in}
}
\maketitle
\begin{abstract}
Uplink power control plays a key role on the performance of uplink cellular network. In this work, the power control factor\footnote{Power control factor ($\in [0,1]$) controls the power transmitted by mobile stations }($\in[0,1]$) is evaluated based on three parameters namely: average transmit power, coverage probability and  average rate. In other words, we  evaluate power control factor such that average transmit power should be low, coverage probability of cell-edge users should be high and also average rate over all the uplink users should be high. We show through numerical studies that the power control factor should be close to $0.5$ in order to achieve an acceptable trade-off between these three parameters.
\end{abstract}
\section{Introduction}
Power control is an important consideration for  the uplink cellular networks. It has two modes of operation: closed loop and open loop power control \cite{uplink}. In closed loop power control, the base station (BS) compare the received  Signal-to-noise-plus-Interference-ratio (SINR) to the desired target SINR. If the received SINR is lesser than the desired target SINR a transmit power control command is transmitted to the mobile station (MS) to increases the transmit power. Otherwise, transmit power control command is transmitted  to decrease the transmit power. On the other hand in open loop power control power there is no feedback path. In this work, an open loop power control is considered.  

Power control factor ($\in [0,1]$) controls the power transmitted by MS. A lower power control factor allows the cell-centre users (users close to BS) to transmit higher power which result higher rate but it also provides higher interference power for the cell-edge users (users far from BS). On the other hand, a higher power control factor allows the cell-centre users  to transmit lower power which result lower rate but it  provides lower interference power for the cell-edge users. Therefore, how should a cellular network operator should choose the power control factor. This is the question we attempt to answer in this work.

The difference between the performance of pure open loop power control and combined open loop and closed loop power control has been studied in \cite{5198853, mullner2009performance}. It has been shown in \cite{5378852} that the fractional path loss compensation factor with closed loop power control can greatly improve the system performance. The impact of fractional power control on the SINR and  interference distribution has been studied in \cite{4526110} and also a sub-optimal configuration is proposed for the fractional power control. It has been shown in \cite{4657149} that fractional path loss compensation is advantageous than the full path loss compensation in terms of cell-edge capacity and also battery life time. A modified fractional power control utilizing the path loss difference between serving cell and strongest interfering cell is proposed in \cite{4350036} to improve the cell-edge bitrate and overall spectral efficiency. Recently, authors of \cite{5876464, 6516885} proposed an analytical approach to this problem and they have provided the insight for choosing the fractional power control. However, although insightful design guidelines are provided, they have not provided the specific parameter value which is at the end, the interest of cellular operator.

In this work, we evaluate the power control factor based on the three parameters namely: average transmit power, coverage probability and average rate. We show that the power control factor should be close to $0.5$. Since at power control factor $=0.5$, average transmit power is low, coverage probability of cell-edge users are moderate and also the average rate is moderate.
 \begin{figure}[ht]
 \centering
 \includegraphics[scale=0.34]{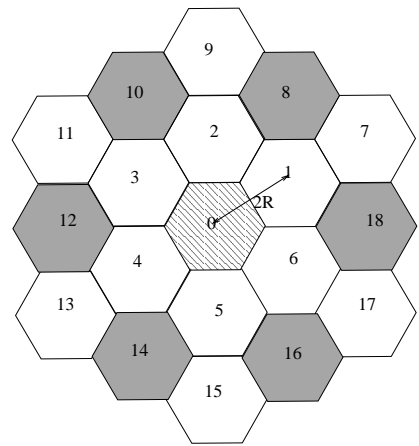}
 \caption{Hexagonal structure of 2-tier macrocell. Interference for $0$th cell in FR$1$ system is contributed form  all the neighbouring $18$ cells, while in a  FR$3$ system it is contributed only from the shaded cells.}
 \label{fig:fig0}
 \end{figure}
\section{System Model}
We consider the uplink  macrocell network having  hexagonal shape  with inter cell site distance $2R$  as shown in Fig. \ref{fig:fig0}. The user is assumed to be uniformly distributed. A path loss model $\l(x)=\|x\|^{-\alpha}$ is considered, where $\alpha \geq 2$ is path loss exponent. Similar to \cite{6516885}, it is assumed that all the MS uses distance-proportion fractional power control factor of the form $d^{\alpha \epsilon}$, where $\epsilon \in [0,1]$ denotes the power control factor. The parameter $\epsilon$ controls the transmit power. $\epsilon =1$ corresponds to an equal received power from all the MS, and at the other end, $\epsilon =0$ corresponds to an equal transmit power.
The received SINR $\eta$ at the tagged BS which is located at origin  from the nearest MS is at the distance $r$ is 
\begin{equation} 
\eta(r)=\frac{gr^{\alpha(\epsilon-1)}}{\sigma^2+I},\text{ where } I=\underset{i \in \phi}\sum h_i d_i^{-\alpha}r_i^{\alpha\epsilon}.
\end{equation}
The distance between the MS to its serving BS (BS at the origin)  is denoted by $r$, and the distances between interfering MSs to their respective serving BSs are denoted by $r_i$. The distance between an interfering MS to the serving BS  at the origin is denoted by $d_i$. We denote the set of  interfering MSs by $\Phi$.
Channel fading gain from tagged BS  and $i^{th}$ interfering MS are denoted by $g$ and $h_i$, respectively, which are independent and identically exponentially distributed with unit mean (corresponding to Rayleigh fading).

\section{Selection of Power Control Factor}
In this  section, we evaluate power control factor based on the three parameters namely: average transmit power, coverage probability, and average rate. The average normalized transmit power used by the MS is given by
\begin{equation}
 P_{\text{avg}}=\int_0^\infty P_t(r)f_R(r)\text{d} r, \label{eq:one1}
\end{equation}
where $P_t(r)$ is the transmit power of MS at distance $r$ from the BS, and $f_R(r)$ is the probability that MS is at distance $r$ from the BS. The probability density function (PDF) of $r$ (and also $r_i$), i.e.,   $f_R(r)$ is given by,
\begin{equation}
f_R(r)= \left\{
\begin{array}{rl}
&\frac{2r}{R^2}, r\leqslant R\\
&0,  r>R.
\end{array} \right.
\end{equation}
Coverage probability is the probability that the measured SINR at the BS of the MS is greater than the target SINR $T$. It is defined as 
\begin{equation}
P_c(T)=\mathbb{P}[\text{SINR}>T].
\end{equation}
The average  rate is calculated based upon the Shannon capacity limit, $R=\mathbb{E}[\ln(1+\eta(r))]$.
\begin{table}[ht]
\caption{Simulation Parameters}
\renewcommand{\tabcolsep}{0.6cm}
\renewcommand{\arraystretch}{2}
\begin{tabular}{|p{4.5cm}|p{1.8cm}|}
\hline 
{\bf Parameters}  & {\bf Values}  \tabularnewline
\hline
\hline 
Carrier Frequency  & 2 GHz \tabularnewline
\hline 
Bandwidth & 10 MHz\tabularnewline
\hline 
Number of Sub-carriers & 600\tabularnewline
\hline 
Number of PRBs & 50\tabularnewline
\hline 
Number of Sub-carriers per PRB  & 12\tabularnewline
\hline 
Total number of users in one macrocell & 34\tabularnewline
\hline 
Number of Interferer cell & 18\tabularnewline
\hline
Macrocell radius ($R$) & 500m \tabularnewline
\hline
$P_{max}^{macro}$ & 43 dBm \tabularnewline
\hline
Noise power &-174 dBm/Hz \tabularnewline
\hline
\end{tabular}
\end{table} 
 \begin{figure}[ht]
 \centering
 \includegraphics[width=\columnwidth]{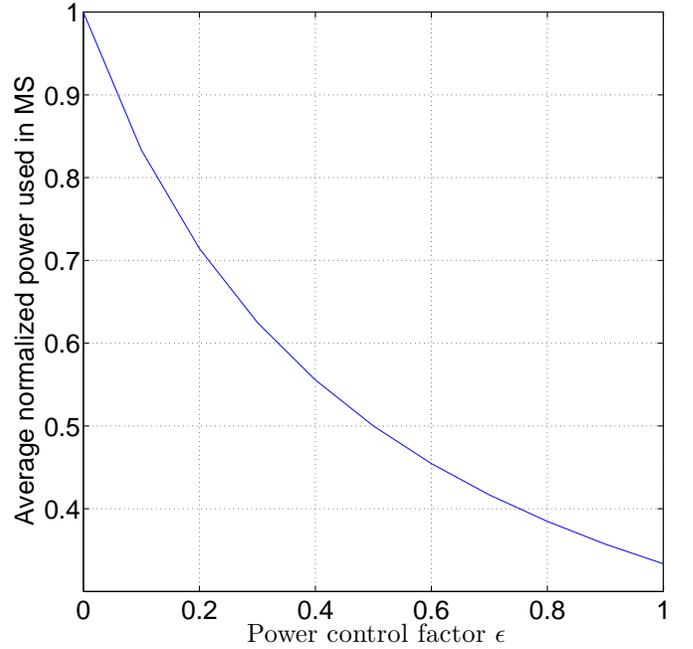}
 \caption{Variation in average power used in the MS with uplink power control factor.}
 \label{fig:fig1}
 \end{figure}
We start the discussion with the average normalized uplink transmit power by the MS. Fig. \ref{fig:fig1} shows the variation in the average normalized power transmitted by the MS versus power control factor using the 
expression given in \eqref{eq:one1}. In can be seen that at $\epsilon = 0$, all the MS will transmit with equal power and hence the average normalized transmitted power is $1$. On the other hand, for $\epsilon=0$, the received power at the BS will be equal and hence the average power transmitted by 
MS  is lowest and it is $0.34$. One important observation  can be made is as follows: as $\epsilon$ increases, the rate of
decreasing average transmitted power  decreases. For example, as $\epsilon$ goes from 0 to 0.1, average power  decreases by $16\%$ 
but at the other extreme, as $\epsilon$ goes from 0.9 to 1, average power  decreases by only $6\%$.

\begin{figure}[ht]
 \centering
 \includegraphics[width=\columnwidth]{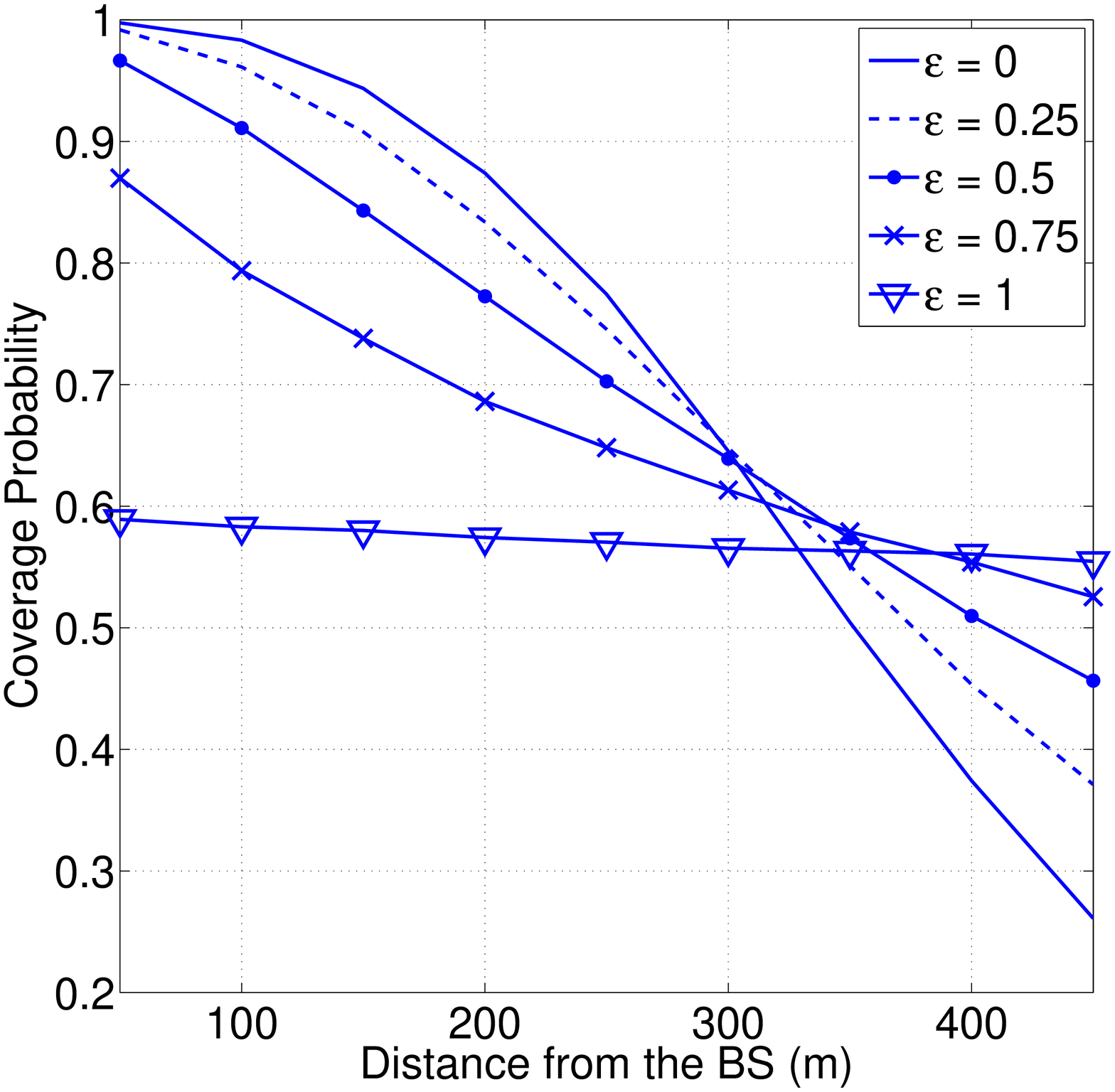}
 \caption{Coverage probability of FR$1$ with respect to distance from the BS for different values of power control factor. Here $\alpha=3$ and $T=0$dB is assumed.}
 \label{fig:fig2}
 \end{figure}

\begin{figure}[ht]
 \centering
 \includegraphics[width=\columnwidth]{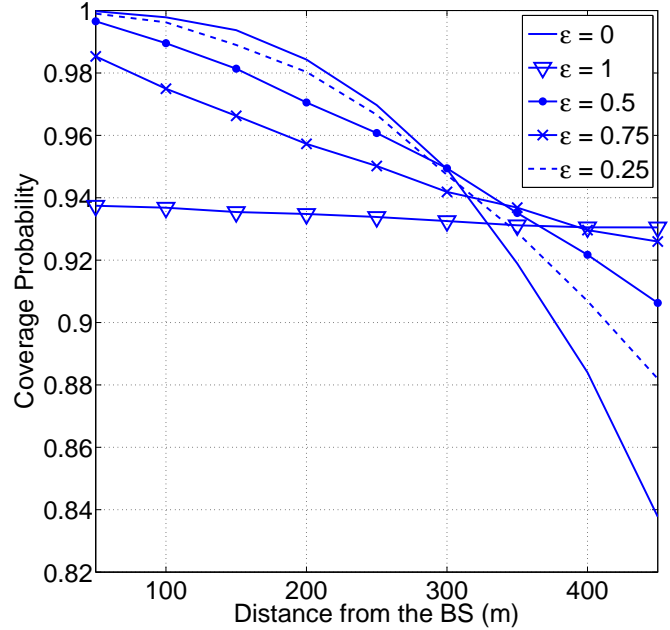}
 \caption{Coverage probability of FR$3$ with respect to distance from the BS for different values of power control factor. Here $\alpha=3$ and $T=0$dB is assumed.}
 \label{fig:fig3}
 \end{figure}

Fig. \ref{fig:fig2} and Fig. \ref{fig:fig3} shows the coverage probability for frequency reuse $1$ (FR$1$) and  frequency reuse $\frac{1}{3}$ (FR$3$) with respect to distance from the BS using the simulation parameters given in Table $1$. The coverage probability is plotted for five different values of power control factor. It is interesting to note that as power control factor increases coverage probability of the cell-centre users  decreases whereas the coverage probability of the cell-edge users increases. It can be also observed that when power control factor increases from $0$ to $0.5$, the coverage probability of cell-centre users does not decrease significantly whereas the coverage probability of cell-edge users increase significantly for both the reuse system.

\begin{figure}[ht]
 \centering
 \includegraphics[width=\columnwidth]{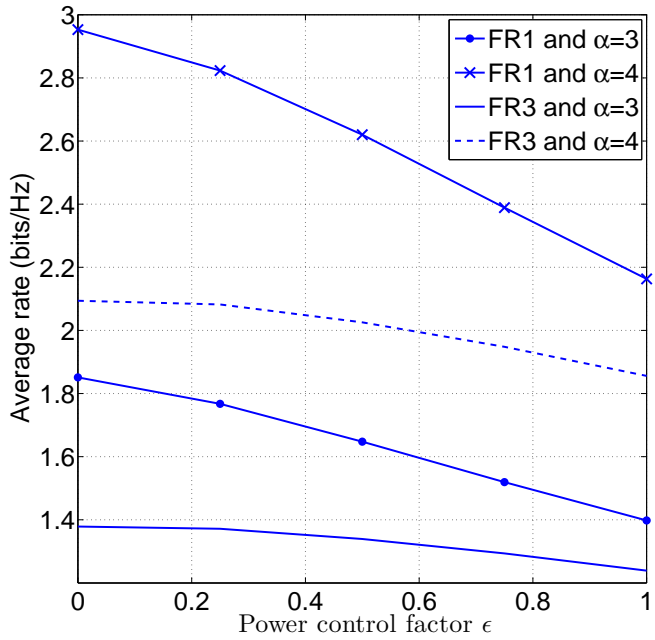}
 \caption{Variation in average rate with respect to power control factor for different reuse schemes and for different path loss exponent.}
 \label{fig:fig4}
 \end{figure}

Fig. \ref{fig:fig4} plots the average rate for the FR$1$ and FR$3$ with respect to power control factor for different path loss exponents. Firstly, it can be observed that as power control factor increases average rate decreases for both the reuse systems and both the path loss exponents. Secondly, it can be observed that as power control factor increase from $0$ to $0.25$ average rate does not decrease significantly (especially for FR$3$ system).

By observing Fig. \ref{fig:fig1}, \ref{fig:fig2}, \ref{fig:fig3} and \ref{fig:fig4}  and behaviour of these plots as discussed  before, one can  conclude that at power control factor $=0.5$, the average transmitted power is decreased by $2$, the coverage probability of cell-edge users increases significantly and also the average rate does not decrease significantly and hence the power control factor should be chosen close to $0.5$.

\begin{figure}[ht]
 \centering
 \includegraphics[width=\columnwidth]{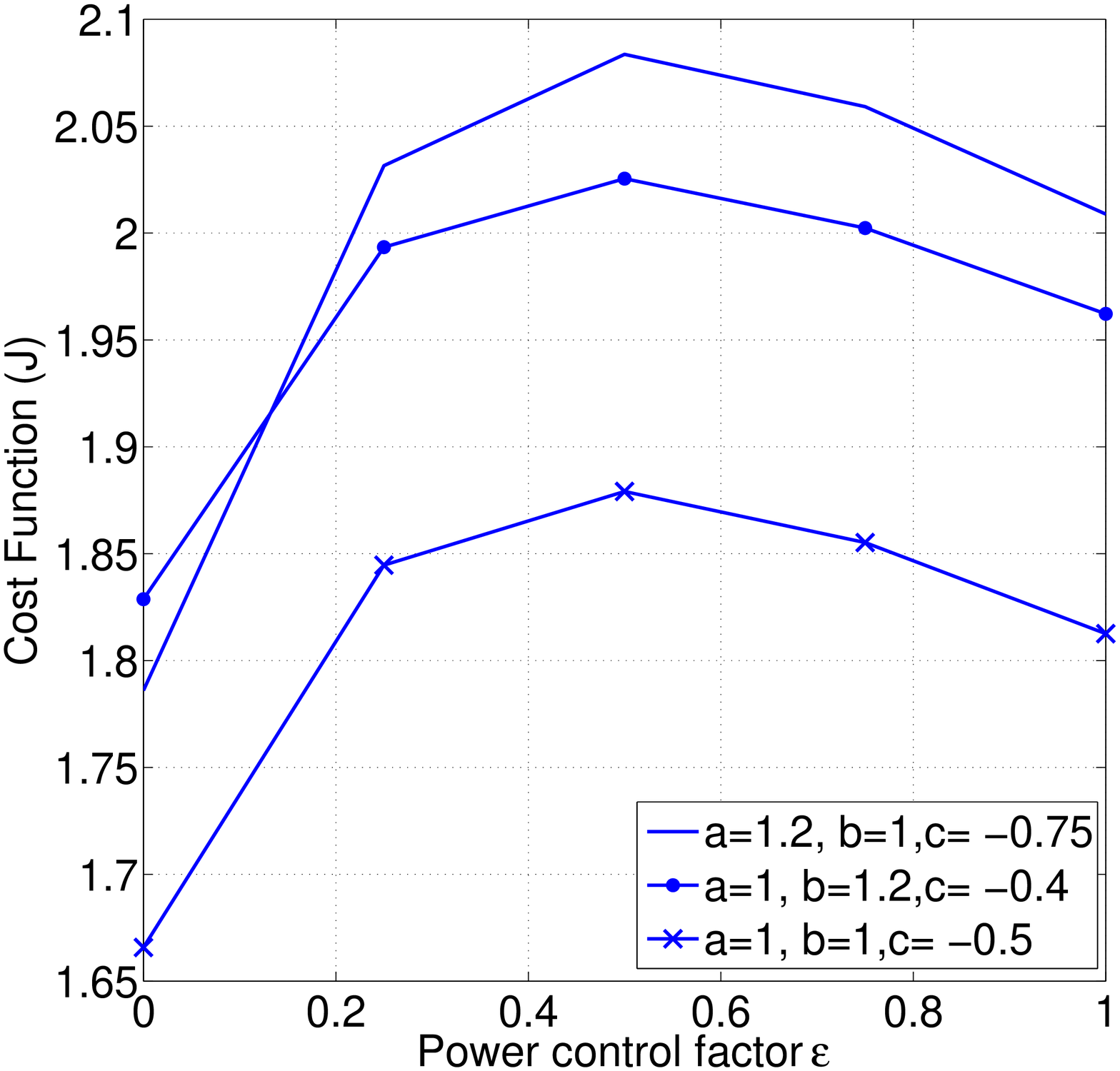}
 \caption{}
 \label{fig:fig5}
 \end{figure}

We have an another way to evaluate the power control factor. We introduce a cost function $J$ which take care of all three parameter: average rate, edge coverage probability and average transmitted power and is given by
\begin{eqnarray*}
J&=&a(\text{average rate})+b(\text{edge coverge probability})\\
&&+ c(\text{average transmitted power}).
\end{eqnarray*}
Here $a>0$, $b>0$ and $c<0$ are weight parameters corresponding to  average rate, edge coverage probability and average transmitted power, respectively. Now, we need to maximize this to evaluate the power control factor. Fig. \ref{fig:fig5} plots the cost function $J$ versus power control factor for three different sets of weight parameter. it can be observe that cost function is maximum at $\epsilon=0.5$. Although, for some sets of parameter cost function will not be maximum at $\epsilon=0.5$, most of the acceptable sets of parameter cost function will be maximum at $\epsilon=0.5$.

\section{Conclusion}
In this paper, we have evaluated the uplink power control factor such that average transmit power should be low, coverage probability of cell-edge users should be high and also average rate should be high. It turns out that power control factor should be close to $0.5$. The natural extension of this work could be to evaluate the uplink power control factor in presence of inter-cell interference coordination  scheme, i.e. fractional frequency reuse \cite{kumar2014optimal}  and  soft frequency reuse \cite{suman}.
\bibliographystyle{IEEEtran}
\bibliography{bibfile}

\end{document}